\documentclass[aps,prd,twocolumn,groupedaddress,showpacs,showkeys,superscriptaddress,nofootinbib]{revtex4}%
\usepackage{graphics}%
\usepackage[mathcal]{euscript}
\usepackage[latin1]{inputenc}
\usepackage{latexsym}
\usepackage{amsmath}
\usepackage{hyperref}
\usepackage{amsmath}    
\usepackage{graphicx}   
\usepackage{verbatim}  
\usepackage{color}      
\usepackage{subfigure}  
\usepackage{hyperref}   
\usepackage{bm}
\usepackage{amssymb}
\usepackage{bbm}
\usepackage{float}
\usepackage{slashed}
\usepackage{amsfonts}
\usepackage{euscript}
\usepackage{amsmath}    
\usepackage{mathrsfs} 
\usepackage{bbding}
\usepackage{pifont}
\usepackage{cancel}

\begin{document}%

\title{Longitudinal diffeomorphisms obstruct the protection of vacuum energy}

\author{Ra\'ul Carballo-Rubio}
\affiliation{Instituto de Astrof\'{\i}sica de Andaluc\'{\i}a (IAA-CSIC), Glorieta de la Astronom\'{\i}a, 18008 Granada, Spain}

\date{12 April 2015}
\bigskip
\begin{abstract}

To guarantee the stability of the cosmological constant sector against radiative corrections coming from quantum matter fields, one of the most natural ingredients to invoke is the symmetry under scale transformations of the gravitational field. Previous attempts to follow this path have nevertheless failed in providing a consistent picture. Here we point out that this failure is intimately tied up to an assumption which is typically embedded in modern studies of the gravitational interaction: invariance under the full group of diffeomorphisms. We base the discussion on the gravitational theory known as Weyl transverse gravity. While leading to the same classical solutions as general relativity, and so to the same classical phenomenology, we show that in the presence of quantum matter (i) the degeneracy between these theories is broken: general relativity exhibits the well-known cosmological constant problem while in Weyl transverse gravity the cosmological constant sector is protected due to gravitational scale invariance, and (ii) this is possible as the result of abandoning the assumption of full diffeomorphism invariance, which permits to circumvent classic results on scale-invariance anomalies and guarantees that gravitational scale invariance survives quantum corrections. Both results signal new directions in the quest of finding an ultraviolet completion of gravity.

\end{abstract}

\pacs{04.20.Cv, 04.60.-m, 04.62.+v}

\keywords{cosmological constant, vacuum energy, scale invariance, emergent gravity, quantum gravity}
\maketitle

\def\e{{\mathrm e}}%
\def\g{{\mbox{\sl g}}}%
\def\Box{\nabla^2}%
\def\d{{\mathrm d}}%
\def\R{{\rm I\!R}}%
\def\ie{{\em i.e.\/}}%
\def\eg{{\em e.g.\/}}%
\def\etc{{\em etc.\/}}%
\def\etal{{\em et al.\/}}%
\def\HRULE{{\bigskip\hrule\bigskip}}

\section{Introduction}

The fact that vacuum zero-point energies of matter do not gravitate can be considered as the only available glance to the realm of quantum gravity with which we have been experimentally rewarded up to now. This view appears precisely because the application of effective field theory arguments to the combination of general relativity and the standard model of particle physics strongly suggests that what is reasonable from the theoretical perspective is, indeed, the contrary \cite{Martin2012}. Finding a mechanism which can forbid these energies to gravitate (or, more formally, to prevent matter radiative corrections to affect the value of the cosmological constant), while keeping intact the low-energy physics as to pass the stringent experimental tests on deviations from these theories, is what is usually known as the cosmological constant problem \cite{Weinberg1989,Burgess2013,Martin2012}.

From an effective field theory perspective, it is well understood that radiative corrections will generate all the terms in the Lagrangian density which are compatible with the symmetries of a given system. In this framework, the cosmological constant term corresponds to a relevant operator that is, however, not natural \cite{Burgess2013,Polchinsk1992}. Therefore the value of the cosmological constant is highly sensitive to the ultraviolet details beyond the effective theory, and it has to be fine-tuned in order to match the experiments. On the one hand, this observation is compelling: the value of the cosmological constant is an issue to be treated in a theory which consistently unifies the ultraviolet and infrared details of our universe, that is, a theory of quantum gravity. On the other hand, it challenges the basic working principle according to which the behavior of physics at a given distance scale is insensitive to the fine details of the dynamics at much shorter distances.

It is therefore desirable that an effective theory rationale exists which permits to work consistently at low energies, leaving the question about the value of the cosmological constant unanswered until we are able to construct a more complete theory. As long as we are concerned only with the gravitational sector, it is straightforward to check that a global symmetry in the form of constant Weyl transformations
\begin{equation}
g_{ab}\rightarrow \zeta^2 g_{ab}\label{eq:protsym}
\end{equation}
with $\zeta$ a real constant would suffice to forbid the occurrence in the effective Lagrangian density of any cosmological constant term. However, when one adds matter one rapidly realizes that there is no way out of the cosmological constant problem by following this path (see, e.g., the discussion in \cite{Burgess2013}): first, only massless fields are allowed in order to preserve the symmetry classically; even in this case, in general the symmetry can no longer be maintained when the matter fields are quantized, with the occurrence of what is known as the Weyl or conformal anomaly \cite{Fujikawa2004,Capper1974,Duff1977}.

In this paper we show that there is a way to embed the symmetry \eqref{eq:protsym} in a consistent effective field theory framework which identifies an alternative gauge structure, distinct from the diffeomorphism group of the spacetime manifold, underneath the classical phenomenology of general relativity. Such a symmetry structure is composed of transverse diffeomorphisms and local Weyl transformations, and to our knowledge it was first discussed in \cite{Izawa1995}. Although different from the symmetry structure usually associated with gravity, it is naturally motivated by the field-theoretical representation and self-coupling problem of gravitons \cite{Alvarez2006,BCRG2014}. In this paper we introduce the geometric conceptualization of these ideas in terms of conformal manifolds.

Even if the different spacetime structure has no effect at the classical level as we will argue below, one may expect differences triggered by quantum effects. We may use the following clear analogy: as radiative corrections can be understood as perturbations with respect to the tree-level physics, this would be equivalent to the (quite common) degeneracy breaking by perturbations in eigenvalue problems. The first of these differences, which affects the renormalization group of the gravitational action in the presence of quantum matter fields, is reported here.

\section{Motivation and geometric construction}

Within the classification of fundamental interactions in terms of the unitary representations of the Poincar\'e group \cite{Wigner1939}, gravity is expected to be mediated by a particle corresponding to the massless spin-2 representation. The massless character of the representation means that only the states with helicity $\pm2$ are physical. This representation has a direct on-shell implementation in field theory \cite{Ortin2004}: a second-rank symmetric, transverse and traceless tensor field $h^{ab}$ which satisfies free equations of motion. In fact, the physical objects are equivalence classes of $h^{ab}$ defined by the equivalence relation of being related through gauge transformations. 

Let us stress that this on-shell description is the sole firm statement one can draw from the assumption that gravity is mediated by a helicity $\pm2$ graviton exclusively, with no admixture of spin 1 or 0. One can take this description as the basis to discuss the self-coupling problem \cite{Barcelo2014}. This procedure has some (tractable) drawbacks which can be avoided by slightly modifying the starting point: the usual way to proceed is to relax the transverse and traceless conditions while enlarging the gauge symmetry, thus maintaining the number of degrees of freedom. This procedure leads to the well-known Fierz-Pauli theory \cite{FierzPauli1939}. However, it is quite remarkable that there exists an alternative extension which also reduces on-shell to a helicity $\pm2$ graviton, which is called Weyl transverse theory \cite{Izawa1995}. As the name suggests, the internal gauge symmetry of the theory is given by
\begin{equation}
{h'}^{ab}\sim h^{ab},\qquad {h'}^{ab}=h^{ab}+\eta^{ac}\partial_c\xi^{b}+\eta^{bc}\partial_c\xi^{a}+\phi\,\eta^{ab},\label{eq:wtsymm}
\end{equation}
with generators satisfying $\partial_a\xi^a=0$, $\phi$ being an arbitrary scalar function, and $\eta^{ab}$ being the Minkowski metric. Fierz-Pauli theory and Weyl transverse theory are the only two embeddings in a linear gauge theory of the on-shell description of gravitons sketched above \cite{Izawa1995,Alvarez2006}. This fact alone is interesting enough to explore this last theory to its ultimate consequences. In principle, even if both are by construction completely equivalent as linear theories, their nonlinear completions could substantially differ. Most importantly, the internal transformations corresponding to the last term in \eqref{eq:wtsymm} can be identified as the local and linear version of a constant Weyl transformation \eqref{eq:protsym}, thus making the discussion of fundamental interest for the cosmological constant problem.

The nonlinear completions of these theories can be motivated by symmetry considerations \cite{Izawa1995,Alvarez2006} or through the self-coupling problem of gravitons \cite{Blas2007,Barcelo2014,BCRG2014}. A detailed analysis shows that these ways are not independent but are intimately related \cite{Barcelo2014}. The nonlinear completion of Fierz-Pauli theory leads to general relativity, or more precisely to Rosen's reformulation of Einstein's theory as a nonlinear field theory over a flat background (see~\cite{Barcelo2014,Ortin2004} for a thorough discussion of the history of the subject and the different assumptions needed to obtain the result). On the other hand, the solution to the self-coupling problem of gravitons which has as non-interacting limit the free Weyl transverse theory is given by the action of Weyl transverse gravity \cite{Alvarez2006,BCRG2014}
\begin{equation}
\mathscr{A}:=\frac{1}{2\kappa}\int\text{d}^Dx\sqrt{|\omega|}\,R[|\omega|^{1/D}|g|^{-1/D}g_{ab}].\label{eq:ren1}
\end{equation}
Here $R[\hat{g}_{ab}]$ has the same functional form as the Ricci scalar of a metric $\hat{g}_{ab}:=|\omega|^{1/D}|g|^{-1/D}g_{ab}$, with $g$ the determinant of $g_{ab}$. We have introduced an auxiliary volume form $\bm{\omega}$ (a nowhere-vanishing differential form of degree $D$), with $\text{d}^Dx\sqrt{|\omega|}$ the corresponding volume element in $D\geq4$ spacetime dimensions, in order to define the action (more details below).

On the other hand, from the perspective of the self-coupling problem $g_{ab}$ is just a tensor field with no a priori metric interpretation. Moreover, the resulting internal symmetries (transverse diffeomorphisms and local Weyl transformations) do not suggest such interpretation, contrary to what happens in the standard construction of general relativity. Although reminiscent of unimodular gravity as presented, e.g., in \cite{Unruh1989,Weinberg1989}, the theory described by the action \eqref{eq:ren1} has as field variable an unconstrained second-rank tensor field $g_{ab}$.

The introduction of the volume form $\bm{\omega}$ is useful to recast the theory in geometric terms and to compare it with the usual formulations of conformal gravity (see, e.g. the recent reviews \cite{Goenner2004,Aalbers2013}). The built-in invariance under Weyl transformations makes conformal manifolds, in which the metric is only defined up to scale, the natural arena to express the theory. The canonical volume form in (pseudo-)Riemannian geometry is not available in a conformal manifold. To find a Weyl-invariant differential form of degree $D$ made up from the gravitational field only, one needs to consider expressions which are at least quartic in the derivatives of the gravitational field, i.e., quadratic in the Weyl tensor \cite{Aalbers2013}. When integrated, this form leads to the usual action of Weyl conformal gravity, which defines a higher-derivative field theory. 

We can dispose however of additional structures compatible with the definition of a differentiable manifold: any orientable manifold naturally admits a space of volume forms \cite{Spivak1971} (for non-orientable manifolds we could equally work with the weaker notion of a density). Although unrelated to the gravitational field, these differential forms can be used to define a useful notion of integration in the conformal manifold (see, e.g., the Appendix B in \cite{Wald2010}). Indeed, using this additional structure permits to write down a second-order action in the derivatives of the gravitational field, thus defining a second-order theory of gravity, in the following way. Given an auxiliary volume form $\bm{\omega}$, construct another differential form of degree $D$ by multiplying it by the scalar curvature $\hat{g}^{cd}R_{cd}(\hat{\bm{\Gamma}})$ of the the Weyl connection $\hat{\bm{\Gamma}}$ \cite{Goenner2004}, whose components are given by
\begin{align}
\hat{\Gamma}^c_{ab}=\frac{1}{2}\hat{g}^{cd}\left(\partial_a\hat{g}_{bd}+\partial_b\hat{g}_{ad}-\partial_d \hat{g}_{ab} \right)=\nonumber\\
=\frac{1}{2}g^{cd}\left(\partial_ag_{bd}+\partial_bg_{ad}-\partial_d g_{ab} \right)+\nonumber\\
+\frac{1}{2D}\left[\delta^c_b\partial_a+\delta^c_a\partial_b-g_{ab}\partial^c\right]\ln(|\omega|/|g|).\label{eq:connect1}
\end{align}
The integral of the differential form constructed in this way leads to the coordinate-free geometric version of the action \eqref{eq:ren1}, 

\begin{equation}
\mathscr{A}=\frac{1}{2\kappa}\int_{\mathscr{M}}\bm{\omega}\,\hat{g}^{ab}R_{ab}(\hat{\bm{\Gamma}}),
\end{equation}
where we have explicitly displayed the domain of integration over the entire spacetime manifold $\mathscr{M}$. This formulation makes explicit the geometric content of the theory: a dynamical conformal structure and an auxiliary volume form.\footnote{It may also suggest to look for ultraviolet completions with a dynamical differential $D-$form, whose dynamics is effectively frozen below certain energy scale.}

\section{Radiative stability}

Matter is included in the discussion by following a minimal coupling approach, but replacing $\eta_{ab}$ with the composite field $\hat{g}_{ab}=|\omega|^{1/D}|g|^{-1/D}g_{ab}$. This can be determined again through symmetry considerations or the self-coupling problem itself. The resulting gravity-matter action is invariant under \emph{gravitational scale transformations}, or local scale transformations of the gravitational field, defined as the local extension of the transformations \eqref{eq:protsym} which \emph{do not affect} matter fields. The space of solutions of such a classical theory is equivalent to the space of solutions of general relativity plus matter, by means of the following argument. The invariance under gravitational scale transformations can be exploited to fix a gauge in which $|g|=|\omega|$. In this gauge the field equations take the same form as the traceless Einstein field equations \cite{BCRG2014,Alvarez2010,Alvarez2012}. These are known to be equivalent to the Einstein field equations, with the cosmological constant appearing as an integration constant \cite{Ellis2011,Ellis2013}. Notice that different choices of the volume form $\bm{\omega}$ are equivalent from the standpoint of the gravitational field equations, as locally these just correspond to picking out different charts on the manifold.

As we discuss in the following, this equivalence breaks up within the semiclassical realm: the radiative corrections to the cosmological constant sector of the theory are identically zero. It is interesting to rephrase the discussion in terms of the Feynman diagrams known as \emph{vacuum bubbles}. These are diagrams with no external legs, so that they represent the (perturbative) view of the quantum vacuum as a `sea' of virtual particles \cite{Milonni1994}. In flat-spacetime quantum field theory, the linked-cluster theorem (see, e.g., \cite{Kopietz2010} for a textbook discussion of the theorem) permits to show explicitly that the contributions of vacuum bubbles cancel out of correlation functions, so that they do not have any physical consequence. If we include gravity in the discussion by means of general relativity and consider the resulting effective quantum field theory \cite{Donogue1994,Burgess2004}, this decoupling no longer holds as a result of the dependence of the spacetime volume form on the gravitational field: diffeomorphism invariance implies the coupling of gravity to these diagrams, leading to the cosmological constant problem \cite{Polchinski2006,Martin2012}.

In the framework of Weyl transverse gravity the corresponding (regulated) radiative corrections to the effective action take the form of a mere constant shift, as we are used to in any special relativistic quantum field theory which \emph{does not} contain gravity. The fact that the spacetime volume form cannot depend on the gravitational field, as discussed above, implies the decoupling of the contributions of vacuum bubbles. In terms of symmetries, we may interpret this result as a consequence of the invariance of the theory under scale transformations of the gravitational field \eqref{eq:protsym}. This symmetry forbids radiative corrections which would otherwise couple to the determinant of the field $g_{ab}$. This can be shown explicitly at one-loop level through, e.g., the effective action scheme \cite{Vassilevich2003,Visser2002,Mukhanov2007}. Given a theory of matter with an arbitrary combination of matter fields with different spin (0, 1/2 and 1), minimally coupled to $\hat{g}_{ab}=|\omega|^{1/D}|g|^{-1/D}g_{ab}$ as we have argued above, the heat kernel expansion \cite{Vassilevich2003} permits to write the regulated effective action in terms of a cutoff $\mu$ so that one can take account of the necessary counterterms.

Let us work this out for a scalar field only, as the generalization to other kinds of matter fields is straightforward. We will follow closely the discussion in \cite{Vassilevich2003,Visser2002}. The evolution equations are given by
\begin{align}
\mathcal{O}_{g_{ab}}\phi=\nonumber\\
=\frac{1}{\sqrt{|\omega|}}\partial_a\left(\sqrt{|\omega|}\,\hat{g}^{ab}\partial_b\phi\right)+m^2\phi+\xi R[\hat{g}_{ab}]\phi=0.\label{eq:scalarop}
\end{align}
The parameters $m$ and $\xi$ are real, but otherwise arbitrary. The first part of the differential operator corresponds to the d'Alembertian operator associated with $\hat{g}_{ab}=|\omega|^{1/D}|g|^{-1/D}g_{ab}$.

Let us now extract the information encoded in the one-loop effective action $\mathscr{S}_{g_{ab}}$, which is essentially the functional determinant of the differential operator $\mathcal{O}_{g_{ab}}$ (a brief discussion of the necessary background is given in the Appendix \ref{sec:app}). Given a fiduciary reference gravitational field $g^0_{ab}$, the following relation is satisfied by the effective action $\mathscr{S}_{g_{ab}}$:
\begin{align}
\mathscr{S}_{g_{ab}}-\mathscr{S}_{g^0_{ab}}=\nonumber\\
=\frac{1}{2}\int\text{d}^4x\,\int_{\mu^{-2}}^\infty\frac{\text{d}s}{s}\left[\exp(-s\,\mathcal{O}_{g^0_{ab}})-\exp(-s\,\mathcal{O}_{g_{ab}})\right].
\end{align}
This expression has been regulated with the introduction of a cutoff $\mu$. Divergences occur when the cutoff is removed, i.e., in the limit $\mu\rightarrow \infty$. To isolate these divergences we can use the heat kernel expansion in the $s\rightarrow0$ limit,
\begin{align}
\exp(-s\,\mathcal{O}_{g_{ab}})=\nonumber\\
=\frac{\sqrt{|\omega|}}{(4\pi s)^2}\left[a_0(\hat{g}_{ab})+a_1(\hat{g}_{ab})s+a_2(\hat{g}_{ab})s^2+\mathscr{O}(s^3)\right].
\end{align}
We have only written explicitly the terms causing divergences in the $s\rightarrow0$ limit, which are associated with the first Seeley-DeWitt coefficients, $\{a_n\}_{n=0,1,2}$. This divergent behavior is absorbed by means of the renormalization of the gravitational couplings. The necessary counterterms to do so can be read from the following expression:
\begin{align}
\mathscr{S}_g=\mathscr{S}_{g_0}-\frac{1}{32\pi^2}\int_{\mathscr{M}}\bm{\omega}\left\{\mu^2[a_1(\hat{g}_{ab})-a_1(\hat{g}^0_{ab})]+\right.\nonumber\\
\left.+\ln(\mu^2/m^2)[a_2(\hat{g}_{ab})-a_2(\hat{g}^0_{ab})]\right\}.\label{eq:hkexp}
\end{align}
The same structure is valid for other kinds of matter fields, just by changing the numeric factors in front of the Seeley-DeWitt coefficients.

From this equation we can notice that there is no term corresponding to the $a_0=1$ Seeley-DeWitt coefficient, in contrast to what happens in general relativity. This is a consequence of the invariance under gravitational scale transformations, which enforces $\sqrt{|\hat{g}|}=\sqrt{|\hat{g}^0|}$ so that these contributions are independent of the gravitational field. The corresponding piece in general relativity leads to the renormalization of the cosmological constant. In general relativity there are also additional terms which renormalize the cosmological constant, coming from the constant pieces (i.e. independent of the gravitational field) of the Seeley-DeWitt coefficients $a_1$ and $a_2$. These automatically cancel out in Weyl transverse gravity, as one can read directly from \eqref{eq:hkexp}. Thus we have shown the main result in this section: in Weyl transverse gravity there is \emph{no} renormalization group equation for the cosmological constant sector.

The first nonzero contribution in \eqref{eq:hkexp} leads to a renormalization of the gravitational coupling constant $\kappa$. If we call $\kappa_0$ the bare gravitational coupling constant then one can read from the one-loop effective action \eqref{eq:hkexp} the equation
\begin{equation}
\frac{1}{\kappa}=\frac{1}{\kappa_0}+C_1\mu^2+C_2\ln\left(\frac{\mu^2}{C_3}\right),\label{eq:reneq}
\end{equation}
where $C_1$, $C_2$ and $C_3$ are constants with convenient physical dimensions and whose values depend on the particle content of the matter sector \cite{Visser2002}. The next contributions involve quadratic expressions in the ``Ricci'' tensor $R_{cd}[\hat{g}_{ab}]$ that also appear in general relativity, which respect gravitational scale invariance and lead to higher-derivative deviations from the second-order field equations at high energies. These imply the running of the corresponding coefficients in front of these quadratic terms in the Lagrangian density, a feature which is irrelevant for the point we want to make here (the form of these renormalization group equations is the same as in the general relativity case and can be consulted, e.g., in \cite{Visser2002}). Again, the important observation is that the invariance of the theory under gravitational scale transformations implies, in contrast with general relativity, the absence of any renormalization group equation describing the running of the cosmological constant sector and, hence, of the corresponding radiative instability. Thus we have one less physically relevant renormalization group equation. To guarantee the consistency of this picture, gravitational scale invariance must survive quantum effects, i.e., be free of anomalies in the presence of quantum matter fields. As we show in the next section, it is perfectly possible to realize this symmetry in a non-anomalous way at the semiclassical level.
 
\section{Absence of anomalies}

Any reader familiar with previous results on scale-invariance anomalies might be skeptical about our statement of protection of the cosmological constant term by means of gravitational scale invariance. While this first reaction is justified, the crucial observation we want to make here is that former results must be revised in the case at hand as one of the previous assumptions has been dropped off: invariance under the entire group of diffeomorphisms. The presence of anomalies can indeed be traced back to this assumption alone, being a feature of the interplay between longitudinal diffeomorphisms and scale transformations. The key resides in the extension of the symmetry \eqref{eq:protsym} when we include matter in the game. The absence of anomalies in Weyl transverse gravity was conjectured in \cite{Blas2008}. The issue was partially studied in \cite{Alvarez2013}, in which the cases of pure gravity, and couplings to \emph{conformal} matter, were considered. These results are unsatisfactorily restricted; for example, only the equation of state $w=1$ would be covered in cosmological solutions. Our results have a broader scope, as they permit to handle the far richer case of a general matter content: arbitrary combinations of fields with different spin (0, 1/2 and 1) and mass parameters, as well as arbitrary interactions between them. In particular, the matter contents of both standard models of particle physics and cosmology, which \emph{are not} conformal, find their place in the present framework.

In the standard case in which it is assumed that the gravitational theory is invariant under diffeomorphisms, the couplings between gravitational and matter fields dictated by symmetry considerations generally imply that matter fields should transform nontrivially under scale transformations, as well as being massless. This results in standard conformal invariance (the situation with scale invariance in quantum chromodynamics is completely parallel). On the contrary, in Weyl transverse gravity the interacting matter field Lagrangian density is constructed from a general flat-space Lagrangian density by replacing the flat metric $\eta_{ab}$ by $\hat{g}_{ab}=|\omega|^{1/D}|g|^{-1/D}g_{ab}$ instead of $g_{ab}$, to guarantee invariance under gravitational scale transformations. Matter fields are inert under these transformations so that no restrictions whatsoever apply. Obviously, this kind of coupling would explicitly break the invariance under longitudinal diffeomorphisms, if present.

To describe anomalies we shall use Fujikawa's approach \cite{Fujikawa1980}, which is especially suited for making our point in a clean and concise way. A specific anomaly will be given by a (regulated) Jacobian associated with the change of the path integral measure under a given symmetry. Even if gravitational scale transformations do not affect matter fields by construction, this symmetry could be anomalous. The best example of this is a scalar field in two dimensions: the fields by itself are unchanged by conformal transformations but still the symmetry is anomalous as the result of the definition of the path integral measure.

For clarity let us briefly recall this well-known two-dimensional conformal invariance example (see, e.g., \cite{Fujikawa2004,Mukhanov2007} and references therein) to compare it with that of gravitational scale invariance (in an arbitrary dimension $D$). We will see that the conformal anomaly arises because of the interplay of conformal transformations and longitudinal diffeomorphisms. That is, the classical conformal symmetry is broken by the quantum anomaly if one uses the path integral measure which preserves invariance under arbitrary diffeomorphisms. For the purposes of this example only, the object $g_{ab}$ will momentarily regain its standard geometric interpretation.

To define the path integral measure and, so, the path integral itself, we first define a scalar product
\begin{equation}
(\phi,\phi'):=\int\text{d}^2x\sqrt{|g|}\,\phi(x)\phi'(x)=\int_\mathscr{M}\bm{\epsilon}\,\phi\,\phi'\label{eq:innprod1}
\end{equation}
in which the differential operator occurring in the classical field equations for the scalar field is symmetric ($\bm{\epsilon}$ is the Levi-Civita tensor \cite{Wald2010}). Then we can perform a decomposition in terms of the eigenfunctions $\{\phi_n\}_{n=1}^\infty$ of this operator. The coefficients of the expansion are given by:
\begin{equation}
b_n:=(\phi_n,\phi)=\int_\mathscr{M}\bm{\epsilon}\,\phi_n\,\phi.\label{eq:ancoeff}
\end{equation}
The classical action can be written entirely in terms of these coefficients. The path integral measure is defined as the natural measure in the infinite-dimensional space spanned by the $\{b_n\}_{n=1}^\infty$, being the path integral formally defined then as a functional determinant. 

By construction, the transformation properties of the measure \label{eq:pimea} are directly inherited from the transformation properties of the inner product \eqref{eq:innprod1} and the fields, and can be directly evaluated from \eqref{eq:ancoeff}. The general result that we will use is the following \cite{Fujikawa2004}: symmetries such that $\delta b_n=0$ are directly free of anomalies, being thus preserved in the quantum realm, while if $\delta b_n\neq0$ additional manipulations are needed in order to extract a meaningful finite result.

The path integral measure constructed from the $\{b_n\}_{n=1}^\infty$ is invariant under diffeomorphisms by construction. Applying the corresponding transformation laws of the fields to the coefficients $b_n$ as defined in \eqref{eq:ancoeff} one can check that these are invariant under these transformations. However, the $\sqrt{|g|}$ factor in the inner product \eqref{eq:innprod1} enforced by diffeomorphism invariance implies that the coefficients \eqref{eq:ancoeff} are not invariant under conformal transformations. Under an infinitesimal conformal transformation $\delta g_{ab}=\alpha(x) g_{ab}$ one has:
\begin{equation}
\delta b_n=\int_\mathscr{M}\bm{\epsilon}\,\phi_n\,\phi\,\alpha.
\end{equation}
It is an algebraic matter to evaluate the Jacobian $J$ associated with the path integral measure for infinitesimal $\alpha(x)$ as:
\begin{equation}
\ln J=\lim_{N\rightarrow\infty}\sum_{n=1}^N\int_\mathscr{M}\bm{\epsilon}\,\phi_n\,\phi_n\,\alpha.
\end{equation}
This expression must be regularized in the $N\rightarrow\infty$ limit but, after a proper treatment, it leads to the usual results of the conformal anomaly: the trace of the stress-energy tensor of matter fields is no longer zero in general gravitational backgrounds so that conformal invariance is lost \cite{Fujikawa2004,Capper1974,Duff1977}.

We can apply the same procedure to show that all the gauge transformations of Weyl transverse gravity are free of anomalies in the presence of quantum matter. To the best of our knowledge, the definition and properties of the path integral for a general matter content coupled to (classical) Weyl transverse gravity have not been discussed before. Again the first we need to do is to define the path integral measure for the matter fields. 

Let us detail the scalar field case, to extend it later to a general matter content. In view of our reasoning in previous sections, the inner product for a scalar field in Weyl transverse gravity will be given by
\begin{equation}
\langle\phi,\phi'\rangle:=\int\text{d}^Dx\sqrt{|\hat{g}|}\,\phi(x)\phi'(x)=\int_{\mathscr{M}}\bm{\omega}\,\phi\,\phi'.\label{eq:inner}
\end{equation}
Notice the essential difference with respect to \eqref{eq:innprod1}, in that this definition displays an auxiliary volume form unrelated to the gravitational field. The inner product is defined for every $\bm{\omega}$ in a coordinate-free way as the integral of the $D-$form $\bm{\omega}\,\phi\,\phi'$ \cite{Wald2010}. 

The evolution operator $\mathcal{O}_{g_{ab}}$ is symmetric in this inner product (see the Appendix \ref{sec:app}). In the following we consider the expansion in terms of its eigenfunctions, which permits to construct the path integral from the coefficients
\begin{align}
c_n:=\langle\phi_n,\phi\rangle=\int_{\mathscr{M}}\bm{\omega}\,\phi_n\,\phi\label{eq:bcoeff}
\end{align}
as:
\begin{align}
\int_{-\infty}^\infty \prod_{n=0}^\infty\frac{\text{d}c_n}{\sqrt{2\pi}}\exp\left(-\sum_{m=1}^\infty\lambda_m c_m^2/2\right)=\mbox{det}^{-1/2}(\mathcal{O}_{g_{ab}}).\label{eq:pathint}
\end{align}
The combination $\sum_{m=1}^\infty\lambda_m c_m^2/2$ corresponds to the classical action, which is invariant by construction under the symmetries of the theory.  

On the other hand,  the transformation properties of the path integral measure can be read off from \eqref{eq:bcoeff}. We can explicitly check that these coefficients and, hence, the corresponding path integral measure, are invariant under both transverse diffeomorphisms and gravitational scale transformations. It is enough to notice that under these transformations $\delta\sqrt{|\hat{g}|}=0$ (or, equivalently, that $\bm{\omega}$ is invariant), thus implying
\begin{equation}
\delta c_n=\int\text{d}^Dx\,\phi_n(x)\phi(x)\,\delta\sqrt{|\hat{g}|}=0.
\end{equation}
In contraposition with the former case we see that, as we are not demanding invariance under longitudinal diffeomorphisms, the factor $\sqrt{|g|}$ can be freely tuned to accommodate the remaining gauge symmetries, generated by gravitational scale transformations. This is what we are effectively realizing when inserting $\hat{g}_{ab}=|\omega|^{1/D}|g|^{-1/D}g_{ab}$ instead of $g_{ab}$ in the expressions for the inner products. These considerations can be extended to \emph{any} matter content, with fields of arbitrary spins and masses, as we can always construct an inner product with these invariance properties as long as we leave aside longitudinal diffeomorphisms and make use of the usual recipes but with the gravitational field represented through the combination $|\omega|^{1/D}|g|^{-1/D}g_{ab}$.

A complementary way to realize this is the following. It is a general result that anomalies may occur in the one-loop effective action when the differential operator appearing in the equations of motions for the fields is not \emph{invariant} under a symmetry, but only \emph{covariant} \cite{Vassilevich2003}. This description is of course parallel to the previous discussion in terms of the measure (a non-invariant operator would imply a non-invariant measure in the path integral). Diffeomorphism invariance enforces a coupling to the gravitational field $g_{ab}$ which implies that it is not possible to make the differential operator invariant under conformal transformations at the same time. On the contrary, when coupling the matter fields to $\hat{g}_{ab}=|\omega|^{1/D}|g|^{-1/D}g_{ab}$ we are giving up invariance of the field equations under longitudinal diffeomorphisms, while making the corresponding differential operators invariant under gravitational scale transformations. An example is the scalar field differential operator defined in \eqref{eq:scalarop}, which is clearly invariant under gravitational scale transformations. Let us illustrate how this works with a different kind of matter field, e.g., a fermion field. It is straightforward to consider more complicated matter contents, such as the one in the Standard Model of Particle Physics, following the recipes for the coupling to the gravitational field given above. Fermions couple to the composite vierbein field, defined by means of the relation $\hat{g}_{ab}=\hat{e}_{\ a}^I\hat{e}_{\ b}^J\eta_{IJ}$ or, in terms of the usual vierbein field,
\begin{equation}
\hat{e}^I_{\ a}=\frac{|\omega|^{1/2D}}{|e|^{1/D}}e^I_{\ a},
\end{equation}
with $|e|$ the determinant of $e^I_{\ a}$. The Dirac operator in Weyl transverse gravity will be given then by
\begin{equation}
\mathcal{D}_{g_{ab}}=i\gamma_I\hat{e}_{\ a}^ID^a-M.\label{eq:dirac}
\end{equation}
Here $M$ is the fermion mass and $D_a$ the covariant derivative associated with $\hat{e}^I_{\ a}$ by means of the corresponding spin connection (see, e.g., \cite{Rovelli2004}), which in our case is given in terms of the Weyl connection \eqref{eq:connect1} as 
\begin{equation}
\hat{\omega}^{IJ}_a=\hat{e}^I_{\ b}\partial_a \hat{e}^{J\,b}-\hat{e}^{I\,b}\hat{e}^J_{\ c}\hat{\Gamma}^c_{ab}.
\end{equation}
While its counterpart in general relativity is just covariant under conformal transformations, the Dirac operator \eqref{eq:dirac} is invariant under gravitational scale transformations, as we have claimed.

Up to now we have covered in the discussion particles with spin 0, 1/2 and 1, with general properties and interactions between them. Let us make a remark concerning the spin-2 case which, although not essential for our semiclassical discussion, may be interesting for future developments. When it comes to the quantum properties of gravity itself, the definition of the path integral is subtler but a tentative exploration of the path integral measure following previous works (see, e.g., \cite{Mottola1995} and references therein) shows that it should be possible to define it in terms of the composite field $\hat{g}_{ab}=|\omega|^{1/D}|g|^{-1/D}g_{ab}$ instead of $g_{ab}$. This procedure would lead to a non-anomalous path integral with respect to the internal symmetries (see also the related discussion for the pure gravity case in \cite{Alvarez2013}). Beyond the scope of the present work, it would be very interesting to carry out this program in detail as well as a study of the possible properties of a quantum theory of gravity with these symmetries.

\section{Scope and prospects}

In the present work we have shown that Weyl transverse gravity is the first known example in the literature of what we can call ``minimal'' solution to the cosmological constant problem: the classical field equations are essentially equivalent to those of general relativity, while the cosmological constant can take arbitrary but radiatively stable values as it is protected by symmetries. Modifications with respect to the classical predictions of general relativity are only triggered by quantum effects, so that tree-level physics is preserved while one-loop and further corrections are changed. It is instructive to observe that the criteria demanded in \cite{Burgess2013} for a satisfactory solution to the cosmological constant problem are verified. From the perspective of the low-energy physics, the cosmological constant is in this framework as mysterious as (but not more than) any other parameter in physics, such as the gravitational constant or the electron charge.

Additionally, this nonlinear gravitational theory has a strong first-principles justification rooted in local particle-like quantum properties of the gravitational interaction, as well as a clean geometric interpretation. This makes it especially suited to describe the infrared limit of a would-be theory of quantum gravity. The construction is, in this sense, inherently different to proposals such as \cite{Kaloper2014} which modify the purely global properties of the gravitational interaction instead of its local properties. From a more philosophical perspective our proposal is inextricably tied up to nontrivial conceptual implications, as one needs to accept, at least at high energies, that some properties of spacetime deviate from the ones associated with a pseudo-Riemannian manifold. This suggests that solving the cosmological constant problem may entail changing our conceptual and mathematical picture of spacetime.

As of future work, from this point different roads can be taken. It would be interesting to explore additional quantum properties of Weyl transverse gravity as an effective theory, apart from the one which is the main subject of this paper and which is in accordance with current observations, in order to further distinguish it from general relativity. Promising candidates to display differences may be scattering amplitudes involving the off-shell structure of gravitons, i.e., containing graviton loops. Regarding the cosmological constant, only a more fundamental theory or principle could unveil its nature and set its actual value, which by matching should be the one used in the classical solutions of the effective low-energy theory. The exploration of the nature of such a principle, which may well be related to the true vacuum of the underlying high-energy theory (see, e.g., Volovik's proposal \cite{Volovik2009}), is an interesting issue in itself. Also the possible construction of theories of quantum gravity with non-anomalous gravitational scale invariance and their relation with the more familiar conformal field theories, appears as an attractive problem which could lead to profound implications for our understanding of the gravitational interaction and scale invariance.

\appendix
\section{Path integral for a scalar field \label{sec:app}}

For the sake of completeness, let us describe some properties of the path integral for a scalar field in the framework of Weyl transverse gravity. Following the usual practice, when performing manipulations with path integrals some expressions should be understood in the Euclidean sense, in order to guarantee that all is well defined (the reader can read the details on this as well as the standard conventions we follow in, e.g., \cite{Bertlmann2000}).

The action leading to the equations of motion \eqref{eq:scalarop} is given by
\begin{equation}
S[\phi,g_{ab}]=\frac{1}{2}\langle\phi,\mathcal{O}_{g_{ab}}\phi\rangle=\frac{1}{2}\int_{\mathscr{M}}\bm{\omega}\,\phi\,\mathcal{O}_{g_{ab}}\phi.\label{eq:appact}
\end{equation}
Integrating by parts twice we can show that the operator $\mathcal{O}_{g_{ab}}$ is symmetric in the inner product \eqref{eq:inner}:
\begin{align}
\langle\phi,\mathcal{O}_{g_{ab}}\phi'\rangle=\nonumber\\
=\int\text{d}^Dx\,\phi\left[\partial_a\left(\sqrt{|\omega|}\,\hat{g}^{ab}\partial_b\phi'\right)+m^2\phi'+\xi R[\hat{g}_{ab}]\phi'\right]=\nonumber\\
=\int\text{d}^Dx\,\phi'\left[\partial_a\left(\sqrt{|\omega|}\,\hat{g}^{ab}\partial_b\phi\right)+m^2\phi+\xi R[\hat{g}_{ab}]\phi\right]=\nonumber\\
=\langle\mathcal{O}_{g_{ab}}\phi,\phi'\rangle.
\end{align}

The corresponding expansion in eigenfunctions permits to write the action as
\begin{equation}
S[\phi,g_{ab}]=\frac{1}{2}\langle\phi,\mathcal{O}_{g_{ab}}\phi\rangle=\frac{1}{2}\sum_{m=1}^\infty\lambda_m c_m^2,
\end{equation}
where $\{c_m\}_{m=1}^\infty$ are the coefficients of the expansion and $\{\lambda_m\}_{m=1}^\infty$ the eigenvalues of the corresponding eigenfunctions. Following the usual normalization conventions the measure is defined in terms of the $\{c_m\}_{m=1}^\infty$ as
\begin{equation}
[\mathscr{D}\phi]:=\prod_{n=0}^\infty\frac{\text{d}c_n}{\sqrt{2\pi}},
\end{equation}
so that the path integral is formally given by
\begin{align}
\int[\mathscr{D}\phi]\exp(-S[\phi,g_{ab}])=\nonumber\\
=\int_{-\infty}^\infty\prod_{n=0}^\infty\frac{\text{d}c_n}{\sqrt{2\pi}}\exp\left(-\sum_{m=1}^\infty\lambda_m c_m^2/2\right)=\nonumber\\
=\prod_{n=0}^\infty\int_{-\infty}^\infty\frac{\text{d}c_n}{\sqrt{2\pi}}\exp\left(-\lambda_n c_n^2/2\right)=\nonumber\\
=\prod_{n=0}^\infty\lambda_n^{-1/2}=\mbox{det}^{-1/2}(\mathcal{O}_{g_{ab}}).
\end{align}
The one-loop effective action is defined through $\exp(-\mathscr{S}_{g_{ab}}):=\int[\mathscr{D}\phi]\exp(-S[\phi,g_{ab}])$, i.e.,
\begin{equation}
\mathscr{S}_{g_{ab}}=\frac{1}{2}\ln\mbox{det}(\mathcal{O}_{g_{ab}}).
\end{equation}
The following definition of the real logarithm
\begin{equation}
\ln(x)=-\lim_{\epsilon\rightarrow0}\int_\epsilon^\infty\frac{\text{d}s}{s}\left[\exp(-xs)-\exp(-s)\right]
\end{equation}
is useful in order to express the effective action in a different way. For two real numbers $\alpha,\beta\in\mathbb{R}$ we have
\begin{equation}
\ln(\alpha/\beta)=\lim_{\epsilon\rightarrow0}\int_\epsilon^\infty\frac{\text{d}s}{s}\left[\exp(-s\beta)-\exp(-s\alpha)\right].
\end{equation}
If we understand these $\alpha$, $\beta$ as eigenvalues of the operators $\mathcal{O}_{g_{ab}}$, $\mathcal{O}_{g^0_{ab}}$ corresponding to two different gravitational fields, and use $\ln\mbox{det}(\mathcal{O}_{g_{ab}})=\mbox{Tr}\ln(\mathcal{O}_{g_{ab}})$ with $\mbox{Tr}$ understood as integration over the coordinates, we can write then
\begin{align}
\mathscr{S}_{g_{ab}}-\mathscr{S}_{g^0_{ab}}=\nonumber\\
=\frac{1}{2}\int\text{d}^4x\,\ln\left(\mathcal{O}_{g_{ab}}/\mathcal{O}_{g^0_{ab}}\right)=\nonumber\\
=\frac{1}{2}\lim_{\epsilon\rightarrow0}\int\text{d}^4x\,\int_{\epsilon}^\infty\frac{\text{d}s}{s}\left[\exp(-s\,\mathcal{O}_{g^0_{ab}})-\exp(-s\,\mathcal{O}_{g_{ab}})\right].
\end{align}
From the way this relation is obtained it is clear that it will be also satisfied for other kinds of matter fields, but in terms of the corresponding differential operators.

Lastly, let us address the robustness of semiclassical physics with respect to the choice of volume form. From, e.g., the expression of the coefficients \eqref{eq:bcoeff}, one may think that given two different volume forms the corresponding semiclassical theories will differ. This intuition comes from thinking about these coefficients as evaluated in a fixed spacetime structure, but the equivalence arises when the dynamical properties of this structure are taken into account, so that the gravitational field configuration is determined by means of the field equations. A change of coordinates suffices to make the field equations for two volume forms $\bm{\omega}$ and $\bm{\omega}'$ formally equivalent, modulo the renaming of the coordinates. As the coordinates are dummy variables the coefficients \eqref{eq:bcoeff} are then invariant. To be quite specific, when picking a specific solution to these equations one has:
\begin{align}
\int_\mathscr{M}\bm{\omega}\,\phi_n\,\phi=\nonumber\\
=\int\text{d}^Dx|\omega|^{1/2}(x)\phi_n(x)\phi(x)=\nonumber\\
=\int\text{d}^Dx'|\omega|^{1/2}(x')\phi_n(x')\phi(x')=\nonumber\\
=\int\text{d}^Dx|\omega'|^{1/2}(x)\phi_n(x)\phi(x)=\nonumber\\
=\int_\mathscr{M}\bm{\omega}'\,\phi_n\,\phi.
\end{align}
This demonstrates that the coefficients \eqref{eq:bcoeff} are independent of the choice of volume form. As the same is true for the eigenvalues of the scalar field differential operator \eqref{eq:scalarop}, being them coordinate invariants, the entire path integral \eqref{eq:pathint} is insensitive to the specific choice of volume form.

\acknowledgments
I wish to thank Carlos Barcel\'o and Luis J. Garay for enlightening discussions and comments on a draft of the paper, and also Miguel Gonz\'alez-Pinto and Jos\'e Alberto Orejuela for reading the manuscript. Financial support was provided by the Spanish MICINN through Project No. FIS2011-30145-C03-01 (with FEDER contribution), by the Junta de Andaluc\'{i}a through Project No. FQM219, and by CSIC through the JAE-predoc program, cofunded by FSE. 

\vskip-15pt
\enlargethispage{75pt}

\bibliography{ldiff_rev}	

\begin{thebibliography}{40}
\expandafter\ifx\csname natexlab\endcsname\relax\def\natexlab#1{#1}\fi
\expandafter\ifx\csname bibnamefont\endcsname\relax
  \def\bibnamefont#1{#1}\fi
\expandafter\ifx\csname bibfnamefont\endcsname\relax
  \def\bibfnamefont#1{#1}\fi
\expandafter\ifx\csname citenamefont\endcsname\relax
  \def\citenamefont#1{#1}\fi
\expandafter\ifx\csname url\endcsname\relax
  \def\url#1{\texttt{#1}}\fi
\expandafter\ifx\csname urlprefix\endcsname\relax\def\urlprefix{URL }\fi
\providecommand{\bibinfo}[2]{#2}
\providecommand{\eprint}[2][]{\url{#2}}

\bibitem[{\citenamefont{{Martin}}(2012)}]{Martin2012}
\bibinfo{author}{\bibfnamefont{J.}~\bibnamefont{{Martin}}},
  \bibinfo{journal}{Comptes Rendus Physique} \textbf{\bibinfo{volume}{13}},
  \bibinfo{pages}{566} (\bibinfo{year}{2012}), \eprint{1205.3365},
  \urlprefix\url{http://arxiv.org/abs/1205.3365}.

\bibitem[{\citenamefont{Weinberg}(1989)}]{Weinberg1989}
\bibinfo{author}{\bibfnamefont{S.}~\bibnamefont{Weinberg}},
  \bibinfo{journal}{Rev. Mod. Phys.} \textbf{\bibinfo{volume}{61}},
  \bibinfo{pages}{1} (\bibinfo{year}{1989}),
  \urlprefix\url{http://link.aps.org/doi/10.1103/RevModPhys.61.1}.

\bibitem[{\citenamefont{{Burgess}}(2013)}]{Burgess2013}
\bibinfo{author}{\bibfnamefont{C.~P.} \bibnamefont{{Burgess}}},
  \bibinfo{journal}{ArXiv e-prints}  (\bibinfo{year}{2013}),
  \eprint{1309.4133}, \urlprefix\url{http://arxiv.org/abs/1309.4133}.

\bibitem[{\citenamefont{{Polchinski}}(1992)}]{Polchinsk1992}
\bibinfo{author}{\bibfnamefont{J.}~\bibnamefont{{Polchinski}}},
  \bibinfo{journal}{ArXiv e-prints}  (\bibinfo{year}{1992}),
  \eprint{hep-th/9210046}, \urlprefix\url{http://arxiv.org/abs/hep-th/9210046}.

\bibitem[{\citenamefont{Fujikawa and Suzuki}(2004)}]{Fujikawa2004}
\bibinfo{author}{\bibfnamefont{K.}~\bibnamefont{Fujikawa}} \bibnamefont{and}
  \bibinfo{author}{\bibfnamefont{H.}~\bibnamefont{Suzuki}},
  \emph{\bibinfo{title}{Path Integrals and Quantum Anomalies}}, International
  Series of Monographs on Physics (\bibinfo{publisher}{Clarendon Press},
  \bibinfo{year}{2004}), ISBN \bibinfo{isbn}{9780198529132},
  \urlprefix\url{http://books.google.de/books?id=z1v8CiRb0X8C}.

\bibitem[{\citenamefont{{Capper} and {Duff}}(1974)}]{Capper1974}
\bibinfo{author}{\bibfnamefont{D.~M.} \bibnamefont{{Capper}}} \bibnamefont{and}
  \bibinfo{author}{\bibfnamefont{M.~J.} \bibnamefont{{Duff}}},
  \bibinfo{journal}{Nuovo Cimento A Serie} \textbf{\bibinfo{volume}{23}},
  \bibinfo{pages}{173} (\bibinfo{year}{1974}),
  \urlprefix\url{http://link.springer.com/article/10.1007%2FBF02748300}.

\bibitem[{\citenamefont{{Duff}}(1977)}]{Duff1977}
\bibinfo{author}{\bibfnamefont{M.~J.} \bibnamefont{{Duff}}},
  \bibinfo{journal}{Nuclear Physics B} \textbf{\bibinfo{volume}{125}},
  \bibinfo{pages}{334} (\bibinfo{year}{1977}),
  \urlprefix\url{http://www.sciencedirect.com/science/article/pii/0550321377904102}.

\bibitem[{\citenamefont{{Izawa}}(1995)}]{Izawa1995}
\bibinfo{author}{\bibfnamefont{K.}~\bibnamefont{{Izawa}}},
  \bibinfo{journal}{Progress of Theoretical Physics}
  \textbf{\bibinfo{volume}{93}}, \bibinfo{pages}{615} (\bibinfo{year}{1995}),
  \eprint{hep-th/9410111},
  \urlprefix\url{http://ptp.oxfordjournals.org/content/93/3/615}.

\bibitem[{\citenamefont{{{\'A}lvarez} et~al.}(2006)\citenamefont{{{\'A}lvarez},
  {Blas}, {Garriga}, and {Verdaguer}}}]{Alvarez2006}
\bibinfo{author}{\bibfnamefont{E.}~\bibnamefont{{{\'A}lvarez}}},
  \bibinfo{author}{\bibfnamefont{D.}~\bibnamefont{{Blas}}},
  \bibinfo{author}{\bibfnamefont{J.}~\bibnamefont{{Garriga}}},
  \bibnamefont{and}
  \bibinfo{author}{\bibfnamefont{E.}~\bibnamefont{{Verdaguer}}},
  \bibinfo{journal}{Nuclear Physics B} \textbf{\bibinfo{volume}{756}},
  \bibinfo{pages}{148} (\bibinfo{year}{2006}), \eprint{hep-th/0606019},
  \urlprefix\url{http://www.sciencedirect.com/science/article/pii/S0550321306006614}.

\bibitem[{\citenamefont{{Barcel{\'o}} et~al.}(2014)\citenamefont{{Barcel{\'o}},
  {Carballo-Rubio}, and {Garay}}}]{BCRG2014}
\bibinfo{author}{\bibfnamefont{C.}~\bibnamefont{{Barcel{\'o}}}},
  \bibinfo{author}{\bibfnamefont{R.}~\bibnamefont{{Carballo-Rubio}}},
  \bibnamefont{and} \bibinfo{author}{\bibfnamefont{L.~J.}
  \bibnamefont{{Garay}}}, \bibinfo{journal}{ArXiv e-prints}
  (\bibinfo{year}{2014}), \eprint{1406.7713},
  \urlprefix\url{http://arxiv.org/abs/1406.7713}.

\bibitem[{\citenamefont{Wigner}(1939)}]{Wigner1939}
\bibinfo{author}{\bibfnamefont{E.}~\bibnamefont{Wigner}},
  \bibinfo{journal}{Annals of Mathematics} \textbf{\bibinfo{volume}{40}},
  \bibinfo{pages}{pp. 149} (\bibinfo{year}{1939}), ISSN
  \bibinfo{issn}{0003486X},
  \urlprefix\url{http://www.jstor.org/stable/1968551}.

\bibitem[{\citenamefont{Ort{\'\i}n}(2007)}]{Ortin2004}
\bibinfo{author}{\bibfnamefont{T.}~\bibnamefont{Ort{\'\i}n}},
  \emph{\bibinfo{title}{Gravity and Strings}}, Cambridge Monographs on
  Mathematical Physics (\bibinfo{publisher}{Cambridge University Press},
  \bibinfo{year}{2007}), ISBN \bibinfo{isbn}{9780521035460},
  \urlprefix\url{http://books.google.es/books?id=HDmucsxABzYC}.

\bibitem[{\citenamefont{Barcel\'o et~al.}(2014)\citenamefont{Barcel\'o,
  Carballo-Rubio, and Garay}}]{Barcelo2014}
\bibinfo{author}{\bibfnamefont{C.}~\bibnamefont{Barcel\'o}},
  \bibinfo{author}{\bibfnamefont{R.}~\bibnamefont{Carballo-Rubio}},
  \bibnamefont{and} \bibinfo{author}{\bibfnamefont{L.~J.} \bibnamefont{Garay}},
  \bibinfo{journal}{Phys.Rev.} \textbf{\bibinfo{volume}{D89}},
  \bibinfo{pages}{124019} (\bibinfo{year}{2014}), \eprint{1401.2941},
  \urlprefix\url{http://journals.aps.org/prd/abstract/10.1103/PhysRevD.89.124019}.

\bibitem[{\citenamefont{{Fierz} and {Pauli}}(1939)}]{FierzPauli1939}
\bibinfo{author}{\bibfnamefont{M.}~\bibnamefont{{Fierz}}} \bibnamefont{and}
  \bibinfo{author}{\bibfnamefont{W.}~\bibnamefont{{Pauli}}},
  \bibinfo{journal}{Royal Society of London Proceedings Series A}
  \textbf{\bibinfo{volume}{173}}, \bibinfo{pages}{211} (\bibinfo{year}{1939}),
  \urlprefix\url{http://rspa.royalsocietypublishing.org/content/173/953/211}.

\bibitem[{\citenamefont{{Blas}}(2007)}]{Blas2007}
\bibinfo{author}{\bibfnamefont{D.}~\bibnamefont{{Blas}}},
  \bibinfo{journal}{Journal of Physics A Mathematical General}
  \textbf{\bibinfo{volume}{40}}, \bibinfo{pages}{6965} (\bibinfo{year}{2007}),
  \eprint{hep-th/0701049},
  \urlprefix\url{http://iopscience.iop.org/1751-8121/40/25/S47}.

\bibitem[{\citenamefont{Unruh}(1989)}]{Unruh1989}
\bibinfo{author}{\bibfnamefont{W.~G.} \bibnamefont{Unruh}},
  \bibinfo{journal}{Phys. Rev. D} \textbf{\bibinfo{volume}{40}},
  \bibinfo{pages}{1048} (\bibinfo{year}{1989}),
  \urlprefix\url{http://link.aps.org/doi/10.1103/PhysRevD.40.1048}.

\bibitem[{\citenamefont{Goenner}(2004)}]{Goenner2004}
\bibinfo{author}{\bibfnamefont{H.}~\bibnamefont{Goenner}},
  \bibinfo{journal}{Living Rev.Rel.} \textbf{\bibinfo{volume}{7}},
  \bibinfo{pages}{2} (\bibinfo{year}{2004}),
  \urlprefix\url{http://relativity.livingreviews.org/Articles/lrr-2004-2/}.

\bibitem[{\citenamefont{Aalbers}(2013)}]{Aalbers2013}
\bibinfo{author}{\bibfnamefont{J.}~\bibnamefont{Aalbers}},
  \emph{\bibinfo{title}{Conformal symmetry in classical gravity}}
  (\bibinfo{year}{2013}),
  \urlprefix\url{http://dspace.library.uu.nl/handle/1874/280136}.

\bibitem[{\citenamefont{Spivak}(1971)}]{Spivak1971}
\bibinfo{author}{\bibfnamefont{M.}~\bibnamefont{Spivak}},
  \emph{\bibinfo{title}{Calculus On Manifolds: A Modern Approach To Classical
  Theorems Of Advanced Calculus}} (\bibinfo{publisher}{Westview Press},
  \bibinfo{year}{1971}), ISBN \bibinfo{isbn}{9780813346120},
  \urlprefix\url{http://books.google.es/books?id=POIJJJcCyUkC}.

\bibitem[{\citenamefont{Wald}(2010)}]{Wald2010}
\bibinfo{author}{\bibfnamefont{R.}~\bibnamefont{Wald}},
  \emph{\bibinfo{title}{General Relativity}} (\bibinfo{publisher}{University of
  Chicago Press}, \bibinfo{year}{2010}), ISBN \bibinfo{isbn}{9780226870373},
  \urlprefix\url{http://books.google.es/books?id=9S-hzg6-moYC}.

\bibitem[{\citenamefont{Alvarez and Vidal}(2010)}]{Alvarez2010}
\bibinfo{author}{\bibfnamefont{E.}~\bibnamefont{Alvarez}} \bibnamefont{and}
  \bibinfo{author}{\bibfnamefont{R.}~\bibnamefont{Vidal}},
  \bibinfo{journal}{Phys.Rev.} \textbf{\bibinfo{volume}{D81}},
  \bibinfo{pages}{084057} (\bibinfo{year}{2010}), \eprint{1001.4458},
  \urlprefix\url{http://journals.aps.org/prd/abstract/10.1103/PhysRevD.81.084057}.

\bibitem[{\citenamefont{Alvarez}(2012)}]{Alvarez2012}
\bibinfo{author}{\bibfnamefont{E.}~\bibnamefont{Alvarez}},
  \bibinfo{journal}{JCAP} \textbf{\bibinfo{volume}{1207}}, \bibinfo{pages}{002}
  (\bibinfo{year}{2012}), \eprint{1204.6162},
  \urlprefix\url{http://iopscience.iop.org/1475-7516/2012/07/002/}.

\bibitem[{\citenamefont{{Ellis} et~al.}(2011)\citenamefont{{Ellis}, {van Elst},
  {Murugan}, and {Uzan}}}]{Ellis2011}
\bibinfo{author}{\bibfnamefont{G.~F.~R.} \bibnamefont{{Ellis}}},
  \bibinfo{author}{\bibfnamefont{H.}~\bibnamefont{{van Elst}}},
  \bibinfo{author}{\bibfnamefont{J.}~\bibnamefont{{Murugan}}},
  \bibnamefont{and} \bibinfo{author}{\bibfnamefont{J.-P.}
  \bibnamefont{{Uzan}}}, \bibinfo{journal}{Classical and Quantum Gravity}
  \textbf{\bibinfo{volume}{28}}, \bibinfo{eid}{225007} (\bibinfo{year}{2011}),
  \eprint{1008.1196},
  \urlprefix\url{http://iopscience.iop.org/0264-9381/28/22/225007/}.

\bibitem[{\citenamefont{Ellis}(2014)}]{Ellis2013}
\bibinfo{author}{\bibfnamefont{G.~F.~R.} \bibnamefont{Ellis}},
  \bibinfo{journal}{Gen.Rel.Grav.} \textbf{\bibinfo{volume}{46}},
  \bibinfo{pages}{1619} (\bibinfo{year}{2014}), \eprint{1306.3021},
  \urlprefix\url{http://link.springer.com/article/10.1007%2Fs10714-013-1619-5}.

\bibitem[{\citenamefont{Milonni}(1994)}]{Milonni1994}
\bibinfo{author}{\bibfnamefont{P.}~\bibnamefont{Milonni}},
  \emph{\bibinfo{title}{The Quantum Vacuum: An Introduction to Quantum
  Electrodynamics}} (\bibinfo{publisher}{Academic Press},
  \bibinfo{year}{1994}), ISBN \bibinfo{isbn}{9780124980808},
  \urlprefix\url{http://books.google.at/books?id=P83vAAAAMAAJ}.

\bibitem[{\citenamefont{Kopietz et~al.}(2010)\citenamefont{Kopietz, Bartosch,
  and Sch{\"u}tz}}]{Kopietz2010}
\bibinfo{author}{\bibfnamefont{P.}~\bibnamefont{Kopietz}},
  \bibinfo{author}{\bibfnamefont{L.}~\bibnamefont{Bartosch}}, \bibnamefont{and}
  \bibinfo{author}{\bibfnamefont{F.}~\bibnamefont{Sch{\"u}tz}},
  \emph{\bibinfo{title}{Introduction to the Functional Renormalization Group}},
  Introduction to the Functional Renormalization Group
  (\bibinfo{publisher}{Springer}, \bibinfo{year}{2010}), ISBN
  \bibinfo{isbn}{9783642050930},
  \urlprefix\url{http://books.google.ca/books?id=cGa5Q9BeMNUC}.

\bibitem[{\citenamefont{{Donoghue}}(1994)}]{Donogue1994}
\bibinfo{author}{\bibfnamefont{J.~F.} \bibnamefont{{Donoghue}}},
  \bibinfo{journal}{\prd} \textbf{\bibinfo{volume}{50}}, \bibinfo{pages}{3874}
  (\bibinfo{year}{1994}), \eprint{gr-qc/9405057},
  \urlprefix\url{http://journals.aps.org/prd/abstract/10.1103/PhysRevD.50.3874}.

\bibitem[{\citenamefont{{Burgess}}(2004)}]{Burgess2004}
\bibinfo{author}{\bibfnamefont{C.~P.} \bibnamefont{{Burgess}}},
  \bibinfo{journal}{Living Reviews in Relativity} \textbf{\bibinfo{volume}{7}},
  \bibinfo{pages}{5} (\bibinfo{year}{2004}), \eprint{gr-qc/0311082},
  \urlprefix\url{http://relativity.livingreviews.org/Articles/lrr-2004-5/}.

\bibitem[{\citenamefont{{Polchinski}}(2006)}]{Polchinski2006}
\bibinfo{author}{\bibfnamefont{J.}~\bibnamefont{{Polchinski}}},
  \bibinfo{journal}{ArXiv High Energy Physics - Theory e-prints}
  (\bibinfo{year}{2006}), \eprint{hep-th/0603249},
  \urlprefix\url{http://arxiv.org/abs/hep-th/0603249}.

\bibitem[{\citenamefont{{Vassilevich}}(2003)}]{Vassilevich2003}
\bibinfo{author}{\bibfnamefont{D.~V.} \bibnamefont{{Vassilevich}}},
  \bibinfo{journal}{Phys. Rep.} \textbf{\bibinfo{volume}{388}},
  \bibinfo{pages}{279} (\bibinfo{year}{2003}), \eprint{hep-th/0306138},
  \urlprefix\url{http://www.sciencedirect.com/science/article/pii/S0370157303003545}.

\bibitem[{\citenamefont{{Visser}}(2002)}]{Visser2002}
\bibinfo{author}{\bibfnamefont{M.}~\bibnamefont{{Visser}}},
  \bibinfo{journal}{Modern Physics Letters A} \textbf{\bibinfo{volume}{17}},
  \bibinfo{pages}{977} (\bibinfo{year}{2002}), \eprint{gr-qc/0204062},
  \urlprefix\url{http://www.worldscientific.com/doi/abs/10.1142/S0217732302006886}.

\bibitem[{\citenamefont{Mukhanov and Winitzki}(2007)}]{Mukhanov2007}
\bibinfo{author}{\bibfnamefont{V.}~\bibnamefont{Mukhanov}} \bibnamefont{and}
  \bibinfo{author}{\bibfnamefont{S.}~\bibnamefont{Winitzki}},
  \emph{\bibinfo{title}{Introduction to Quantum Effects in Gravity}}
  (\bibinfo{publisher}{Cambridge University Press}, \bibinfo{year}{2007}), ISBN
  \bibinfo{isbn}{9780521868341},
  \urlprefix\url{http://books.google.es/books?id=vmwHoxf2958C}.

\bibitem[{\citenamefont{{Blas}}(2008)}]{Blas2008}
\bibinfo{author}{\bibfnamefont{D.}~\bibnamefont{{Blas}}},
  \bibinfo{journal}{ArXiv e-prints}  (\bibinfo{year}{2008}),
  \eprint{0809.3744}, \urlprefix\url{http://arxiv.org/abs/0809.3744}.

\bibitem[{\citenamefont{{{\'A}lvarez} and {Herrero-Valea}}(2013)}]{Alvarez2013}
\bibinfo{author}{\bibfnamefont{E.}~\bibnamefont{{{\'A}lvarez}}}
  \bibnamefont{and}
  \bibinfo{author}{\bibfnamefont{M.}~\bibnamefont{{Herrero-Valea}}},
  \bibinfo{journal}{Phys. Rev. D} \textbf{\bibinfo{volume}{87}},
  \bibinfo{eid}{084054} (\bibinfo{year}{2013}), \eprint{1301.5130},
  \urlprefix\url{http://journals.aps.org/prd/abstract/10.1103/PhysRevD.87.084054}.

\bibitem[{\citenamefont{Fujikawa}(1980)}]{Fujikawa1980}
\bibinfo{author}{\bibfnamefont{K.}~\bibnamefont{Fujikawa}},
  \bibinfo{journal}{Phys. Rev. D} \textbf{\bibinfo{volume}{21}},
  \bibinfo{pages}{2848} (\bibinfo{year}{1980}),
  \urlprefix\url{http://link.aps.org/doi/10.1103/PhysRevD.21.2848}.

\bibitem[{\citenamefont{Rovelli}(2004)}]{Rovelli2004}
\bibinfo{author}{\bibfnamefont{C.}~\bibnamefont{Rovelli}},
  \emph{\bibinfo{title}{Quantum Gravity}}, Cambridge Monographs on Mathematical
  Physics (\bibinfo{publisher}{Cambridge University Press},
  \bibinfo{year}{2004}), ISBN \bibinfo{isbn}{9780521837330},
  \urlprefix\url{http://books.google.es/books?id=HrAzTmXdssQC}.

\bibitem[{\citenamefont{{Mottola}}(1995)}]{Mottola1995}
\bibinfo{author}{\bibfnamefont{E.}~\bibnamefont{{Mottola}}},
  \bibinfo{journal}{Journal of Mathematical Physics}
  \textbf{\bibinfo{volume}{36}}, \bibinfo{pages}{2470} (\bibinfo{year}{1995}),
  \eprint{hep-th/9502109},
  \urlprefix\url{http://scitation.aip.org/content/aip/journal/jmp/36/5/10.1063/1.531359}.

\bibitem[{\citenamefont{{Kaloper} and {Padilla}}(2014)}]{Kaloper2014}
\bibinfo{author}{\bibfnamefont{N.}~\bibnamefont{{Kaloper}}} \bibnamefont{and}
  \bibinfo{author}{\bibfnamefont{A.}~\bibnamefont{{Padilla}}},
  \bibinfo{journal}{Physical Review Letters} \textbf{\bibinfo{volume}{112}},
  \bibinfo{eid}{091304} (\bibinfo{year}{2014}), \eprint{1309.6562},
  \urlprefix\url{http://journals.aps.org/prl/abstract/10.1103/PhysRevLett.112.091304}.

\bibitem[{\citenamefont{Volovik}(2009)}]{Volovik2009}
\bibinfo{author}{\bibfnamefont{G.}~\bibnamefont{Volovik}},
  \emph{\bibinfo{title}{The Universe in a Helium Droplet}}, International
  Series of Monographs on Physics (\bibinfo{publisher}{OUP Oxford},
  \bibinfo{year}{2009}), ISBN \bibinfo{isbn}{9780199564842},
  \urlprefix\url{http://books.google.es/books?id=6uj76kFJOHEC}.

\bibitem[{\citenamefont{Bertlmann}(2000)}]{Bertlmann2000}
\bibinfo{author}{\bibfnamefont{R.}~\bibnamefont{Bertlmann}},
  \emph{\bibinfo{title}{Anomalies in Quantum Field Theory}}, International
  Series of Monographs on Physics (\bibinfo{publisher}{Clarendon Press},
  \bibinfo{year}{2000}), ISBN \bibinfo{isbn}{9780198507628},
  \urlprefix\url{http://books.google.es/books?id=FC\_DRRUHFXEC}.

\end{thebibliography}

\end{document}